# Recent Results on Photoproduction of Vector Mesons and Mass Dependent Pomeron Trajectory

Shaukat Ali and Muhammad Ali

Theory Group, Department of Space Science,

University of the Punjab, Lahore.

**ABSTRACT**

It is shown that the recent results on photoproduction of vector mesons can be naturally explained by using a mass dependet pomeron trajectory.

Regge theory [1] has been successful in describing soft hadron-hadron interactions. Elastic photoproduction of $\rho^0$ and $\varphi$ mesons is well described [2] as a diffractive process in the framework of the Vector Dominance Model (VDM) [3] and Regge theory [1]. Soft diffraction can be described by the exchange of a soft pomeron Regge trajectory $\alpha(t) = \alpha(0) + \alpha' t$ with an intercept $\alpha(0) = 1.08$ and slope $\alpha' = 0.25$ GeV$^{-2}$. This trajectory is also sometimes referred to as the nonperturbative Pomeron. However, the steep energy behaviour of the elastic $J/\Psi$ photoproduction cross section at HERA, cannot be described in the Regge picture by a pomeron trajectory with an intercept of 1.08. The cross section for this reaction rises faster than expected from a 'soft' diffractive reaction. While the cross section of elastic photoproduction of the three lightest vector mesons $\rho^0$, $\omega$ and $\varphi$, rises like $W^{0.22}$ [4] that of the $J/\Psi$ rises like $W^{0.64\pm0.13}$ [5] or $W^{0.92\pm0.14}$ [6]. This indicates that the intercept of the exchanged 'object' in the latter reaction is larger than 1.08 but in order to measure the intercept from the cross section behavior one needs information also on the slope $\alpha'$.

We have argued that [7] the elastic photoproduction of $\rho^0$, $\omega$ and $\varphi$ meson ( soft diffractive photoproduction ) as well as that of the $J/\Psi$ photoproduction ( hard diffractive process) can be described by assuming mass dependent expression for the pomeron trajectory in the frame work of Regge theory [1]. It is interesting that the elastic photoproduction data of $\rho^0$,$\varphi$ and $J/\Psi$ vector meson, recently measured by ZEUS Collaboration [8] up to $-t = 3$ GeV$^2$ and at average centre of mass energy $<W> = 94$ GeV agree well with our fit except $\rho^0$. A careful analysis shows that some other quantity also plays role in these reactions. Thus in order to fit the experimental data, we also have to consider the role of isospin I in the amplitude. We have shown that by proposing the isospin dependent residue function of the form exp (-3It) in the framework of Regge theory, these measurements for all the three reactions along with the previous results at high energies can be explained by using the mass dependent pomeron trajectory.

We suggest a simple expression as

$$\alpha(t) = (1 + fm) + (a/m)t$$

where intercept $\alpha(0) = 1+ fm$ and slope $\alpha' = a/m$ GeV$^{-2}$, $f = 0.05$ GeV$^{-1}$, $a = 0.20$ GeV$^{-1}$ and m is the mass of the vector meson involved in the process $\gamma p \to Vp$. For lighter vector meson $\rho^0$ and $\varphi$ this expression behaves like a soft pomeron trajectory and for elastic J/Ψ photoproduction behaves like a hard pomeron trajectory.

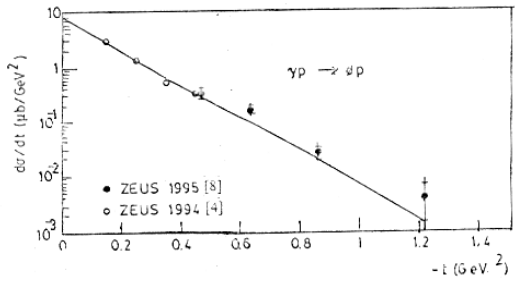 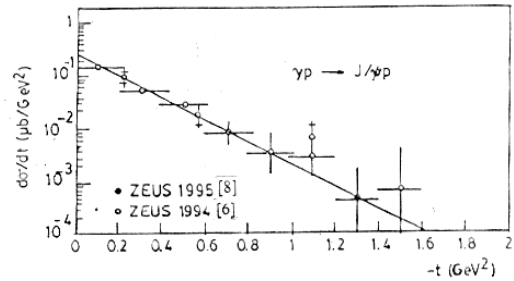

The differential cross-section results for γp→ φp    The differential cross-section results for γp → J/Ψp

By using this simple expression, the differential cross sections of the soft and hard diffractive reactions, γp → Vp, V = ρ$^0$, φ and γp → J/Ψp along with the previous results at high energies are explained. The differential cross-section results for γp → ρ$^0$p and γp →φp reactions calculated by using mass dependent pomeron trajectory are compared with the recently measured experimental data [8]. The agreement is quite good. Next we compare the predictions of our calculations with the experimental data [8] for γp → J/Ψp reaction. The theoretical predictions of the model are in good agreement with the experimental data. The intercept of the exchange pomeron in this reaction is 1.153 and slope is about .07t which is very small as compared to soft pomeron, giving the evidence for no shrinkage. This can be interpreted as indication that for the reaction γp → J/Ψp diffusion from small to large size configuration is only a small correction up to W = 94 GeV. Thus the process is hard and fully calculable in purturbative QCD

ACKNOWLEDGMENT: We are grateful to Professor M. Saleem for drawing our attention to this problem.